\documentclass[fleqn,twoside]{article}%
\usepackage{amssymb}
\usepackage{amsmath}
\usepackage{amsfonts}
\usepackage{espcrc2}
\usepackage{graphicx}%
\providecommand{\U}[1]{\protect\rule{.1in}{.1in}}
\begin{document}

%declarations for front matter%
%TCIMACRO{\TeXButton{Title}{\title{Revisiting the Fermi Golden Rule:
%Quantum Dynamical Phase Transition as a Paradigm Shift}}}%
%BeginExpansion
\title{Revisiting the Fermi Golden Rule:
Quantum Dynamical Phase Transition as a Paradigm Shift}%
%EndExpansion
%

%TCIMACRO{\TeXButton{Author}{\author{Horacio M. Pastawski\address
%{Facultad de Matemática, Astronomía y Física,
%Universidad Nacional de Córdoba, \\
%5000 Córdoba, Argentina}        \thanks{e-mail: horacio@famaf.unc.edu.ar}
%}}}%
%BeginExpansion
\author{Horacio M. Pastawski\address
{Facultad de Matemática, Astronomía y Física,
Universidad Nacional de Córdoba, \\
5000 Córdoba, Argentina}        \thanks{e-mail: horacio@famaf.unc.edu.ar}
}%
%EndExpansion
%

%TCIMACRO{\TeXButton{Begin abstract}{\begin{abstract}}}%
%BeginExpansion
\begin{abstract}%
%EndExpansion
Classical and quantum phase transitions involve observables which are
non-analytic as functions of a controlled thermodynamical variable. As occurs
with the self-consistent Fermi Golden Rule, one condition to obtain the
discontinuous behavior is the proper evaluation of a classical or quantum
thermodynamic limit. We show that\ in presence of an environment, the
oscillatory dynamics of a quantum two-level system, in analogy with a
classical damped oscillator, can undergo a quantum dynamical phase transition
to a non-oscillatory phase. This is obtained from a self-consistent solution
of the Generalized Landauer B\"{u}ttiker Equations, a simplified integral form
of the Keldysh formalism. We argue that working at each side of the transition
implies standing under different paradigms in the Kuhn's sense of the word. In
consequence, paradigms incommensurability obtains a sound mathematical
justification as a consequence of the non-analyticity of the observables. A
strong case is made upon the need to deepen the public's intuition and
understanding on the abrupt transition from static to dynamical friction regimes.

key-words: Paradigm Shift, Quantum Dynamical Phase Transition, Dissipative
Two-Level Systems, Self-Consistent Fermi Golden Rule, thermodynamic limit
%TCIMACRO{\TeXButton{End abstract}{\end{abstract}}}%
%BeginExpansion
\end{abstract}%
%EndExpansion
%

%TCIMACRO{\TeXButton{Make title}{\maketitle}}%
%BeginExpansion
\maketitle
%EndExpansion

\section{Introduction}

From metal and glass melting to steam engines, phase transitions have nurtured
both technological and scientific progress. Only in the last century has it
become clear that phase transitions occur when the relevant free energy is
non-analytic on some controlled thermodynamical variables such as the
temperature. However, it is not trivial to see how this collective behavior
emerges from the fundamental interactions governing microscopic variables.
More recently, quantum phase transitions \cite{Sachdev}, which are much more
elusive, have received an increasing interest. This is mainly motivated by the
High-Tc superconductors, transport in heavy fermion compounds and organic
conductors and condensation of bosonic fluids. The greatest achievement was
perhaps the theory of the extended-localized transition of\ electrons in
disordered solids discovered by P. W. Anderson in 1958 \cite{Anderson NOBEL}.
The whole solid state community was taken by surprise by his statement that
non-interacting electronic or vibrational eigenstates in solids would
transform from Bloch plane waves into exponentially localized functions
whenever the strength of a homogeneous disorder exceeds a critical value.
Typically, one should recognize a phase transition as a non-analytic behavior
of the ground state energy or other observable as a function of a control
parameter $g$~\cite{Sachdev}. This sort of phenomena could \ trivially occur
when the control parameter moves the system through a level crossing. However,
this involves a total Hamiltonian of the form $H_{1}+gH_{2}$ with $H_{1}$ and
$H_{2}$ mutually commuting. In finite systems this would be an extremely rare
situation, but it becomes more likely when one considers an infinite lattice.
In this case, the infinite number of degrees of freedom involved could
transform an avoided crossing of the finite system into an actual level-crossing.

In this work we want to discuss how a quantum dynamics of a system can undergo
a phase transition. We consider a system tunneling coherently between two
levels to form a Rabi oscillation. This system is ubiquitous in Nature
\cite{Feynman-Lect-III}, but has received renewed attention in quantum
information field because it constitutes a swapping gate
\cite{Nakamura-exp-swapp,Myatt-exp-swapp,Alvarez-exp-swapp}. The presence of a
quantum environment, requires the solution of the dynamics of open systems
\cite{spin-boson}. We resort to the Keldysh formalism\cite{Keldysh} which,
with some simplifying assumptions, becomes the Generalized
Landauer-B\"{u}ttiker Equations \cite{GLBE1,GLBE2} which can be solved
analytically. We find that the oscillatory dynamics can freeze when the
interaction with a quantum environment exceeds certain critical strength. This
behavior has a close analogy with the transition between dynamical regimes
(oscillating-overdamped) undergone by a classical oscillator when friction is
increased. Since several of the current descriptions of these phenomena do not
point out the conceptual assumptions enabling the phase transition, in this
article I will sketch out the calculations focusing on the conceptual
conundrums: What is the meaning of a `thermodynamic limit' in classical and
quantum mechanics? Why does the quantum description of an open system involve
a form of thermodynamic limit, and why can this enable a quantum dynamical
phase transition?

Finally, I will conclude with a section associating phase transitions to a
paradigm shift in science \cite{Kuhn-struct}. Similarly to what occurred with
the Aristotelian-Newtonian shift, the mechanicists-probabilistic shift
manifested in the well known Loschmidt vs. Boltzmann polemics (that switches
between reversible and irreversible mechanics) and the related
Zermelo/Poincar\'{e} vs. Boltzmann\ argument on the transition between
recurrent and dissipative mechanics\cite{Kuhn-Blackbody}.

\section{Effective Hamiltonians}

We are particularly interested in the coherent polarization transfer among two
magnetic nuclei, which can be reduced to a non-interacting electron
\cite{spin->fermion} so we will resume the basic formulation of the latter
problem \cite{Pastawski-Medina}. The real symmetric Hamiltonian $\hat{H}%
=\hat{H}^{(0)}+\hat{V}\mathbf{,}$ describes\ the dynamics of two states,
$\left\vert A\right\rangle =\hat{c}_{A}^{+}\left\vert \emptyset\right\rangle
~$and $\left\vert B\right\rangle =\hat{c}_{B}^{+}\left\vert \emptyset
\right\rangle $ which are mixed by a tunneling matrix element $-V_{{\small AB}%
}$. In matrix representation,%
\begin{gather}
\left[  \mathbf{H}^{(0)}+\mathbf{V}\right]  \vec{u}=\varepsilon\mathbf{I}%
\vec{u}\,\ \ \mathrm{with}\label{Ho + V}\\
\mathbf{H}^{(0)}=\left[
\begin{array}
[c]{cc}%
E_{{\small A}} & 0\\
0 & E_{{\small B}}%
\end{array}
\right]  ~\mathrm{and~}\mathbf{V}=\left[
\begin{array}
[c]{cc}%
0 & -V_{{\small AB}}\\
-V_{{\small BA}} & 0
\end{array}
\right]  .\nonumber
\end{gather}
Eliminating one of the amplitudes, e.g. $u_{B},$ gives%
\begin{equation}
\overset{{\LARGE H}_{{\small A}}^{\mathrm{eff.}}}{[\overbrace{E_{{\small A}%
}+\underset{{\LARGE \Sigma}_{{\small A}}}{\underbrace{V_{AB}\dfrac
{1}{\varepsilon-E_{B}}V_{BA}]}}}}u_{{\small A}}=\varepsilon~u_{{\small A}%
}.\label{Ua}%
\end{equation}
Obviously the bracket is an \textquotedblleft effective\textquotedblright%
\ Hamiltonian $H_{A}^{\mathrm{eff.}}=\bar{E}_{{\small A}}(\varepsilon)$ which
includes the \textquotedblleft energy shift\textquotedblright\ $\Sigma
_{{\small A}}(\varepsilon)$ due to the eliminated orbital
\begin{equation}
\bar{E}_{{\small A}}(\varepsilon)=E_{A}+\Sigma_{{\small A}}(\varepsilon
),\label{EeffA}%
\end{equation}%
\begin{equation}
\Sigma_{{\small A}}(\varepsilon)=V_{{\small AB}}\dfrac{1}{(\varepsilon
-E_{{\small B}})}V_{{\small BA}}.\label{SigmaAA}%
\end{equation}
Indeed, under an apparent simplicity, the equation becomes non-linear and it's
solution provides the\ two exact eigenvalues of the system
\begin{align}
\varepsilon_{{\small A}} &  =\tfrac{1}{2}[(E_{{\small A}}+E_{{\small B}%
})-\hbar\omega_{{\small AB}}],\label{Ea}\\
\varepsilon_{B} &  =\tfrac{1}{2}[(E_{{\small A}}+E_{{\small B}})+\hbar
\omega_{{\small AB}}]\label{EB}\\
\hbar\omega_{AB} &  =\sqrt{(E_{B}-E_{{\small A}})^{2}+4\left\vert
V_{AB}\right\vert ^{2}}\label{Wab}%
\end{align}
This procedure can also be expressed in terms of Green's functions. Given a
positive $\eta$, one defines the retarded and advanced resolvent matrices,
\begin{align}
\mathbf{G}^{R}\left(  \varepsilon+\mathrm{i}\eta\right)   &  =\left[
\mathbf{G}^{A}\left(  \varepsilon-\mathrm{i}\eta\right)  \right]  ^{\dagger
}\label{Ga=Gb}\\
&  =\left[  \left(  \varepsilon+\mathrm{i}\eta\right)  \mathbf{I}%
\,-\mathbf{H}\right]  ^{-1}\\
&  =\underset{(\varepsilon+i\eta-\varepsilon_{A})(\varepsilon+i\eta
-\varepsilon_{B})}{\tfrac{1}{\underbrace{\left(  \varepsilon-E_{A}\right)
\left(  \varepsilon+\mathrm{i}\eta-E_{B}\right)  -V_{AB}V_{BA}}}}\nonumber\\
&  \times\left[
\begin{array}
[c]{cc}%
{\small \varepsilon+\mathrm{i}\eta-E}_{{\small B}} & {\small -V}_{{\small AB}%
}\\
{\small -V}_{{\small BA}} & {\small \varepsilon-E}_{{\small A}}%
\end{array}
\right]  .
\end{align}
The retarded (advanced) Green's functions are matrix elements which, for real
$\varepsilon$, have divergences at the eigen-energies as $\eta\rightarrow
0^{+}$ being analytic in the upper (lower) half plane. These divergencies
weigh the probability of the unperturbed state on the eigenstates $\left\vert
\bar{A}\right\rangle $ and $\left\vert \bar{B}\right\rangle .$ Hence, the
local density of states (LDoS) at site $n=A,B$ results:%

\begin{align}
N_{n}(\varepsilon) &  =-\tfrac{1}{\pi}\lim_{\eta\rightarrow0^{+}%
}\operatorname{Im}\left\langle n\right\vert \hat{G}^{oR}(\varepsilon
+\mathrm{i}\eta)\left\vert n\right\rangle \\
&  =-\tfrac{1}{2\pi}\left[  G_{{\small n,n}}^{oR}(\varepsilon)+G_{{\small n,n}%
}^{oA}(\varepsilon)\right]  \nonumber\\
&  =\left\vert \left\langle {\small n}\right\vert \left.  \bar{A}\right\rangle
\right\vert ^{2}\delta(\varepsilon-\varepsilon_{{\small A}})+\left\vert
\left\langle {\small n}\right\vert \left.  {\tiny \bar{B}}\right\rangle
\right\vert ^{2}\delta(\varepsilon-\varepsilon_{{\small B}}).\nonumber
\end{align}

The diagonal matrix elements can be rewritten as
\begin{equation}
G_{A,A}^{R}(\varepsilon)=\frac{1}{\varepsilon-\bar{E}_{{\small A}}%
(\varepsilon)},\label{GAA}%
\end{equation}

Identifying the unperturbed Green's functions $G_{n,n}^{oR}(\varepsilon
)=\left[  \varepsilon-E_{n}\right]  ^{-1}$ and expanding one gets,%

\begin{align}
G_{A,A}^{R}(\varepsilon) &  =\frac{1}{\left[  G_{A,A}^{oR}(\varepsilon
)\right]  ^{-1}-\Sigma_{{\small A}}(\varepsilon)}\nonumber\\
&  =G_{A,A}^{oR}(\varepsilon)+G_{{\small A,A}}^{oR}(\varepsilon)\Sigma
_{A}(\varepsilon)G_{{\small A,A}}^{oR}(\varepsilon)\nonumber\\
&  +G_{{\small A,A}}^{oR}(\varepsilon)\Sigma_{{\small A}}(\varepsilon
)G_{{\small A,A}}^{oR}(\varepsilon)\label{GAA-summ}\\
&  ~~~~~~~~~~~\times\Sigma_{{\small A}}(\varepsilon)G_{{\small A,A}}%
^{oR}(\varepsilon)+....\nonumber
\end{align}
This shows that the exact solution is the sum of an infinite geometric series.
This is represented as Feynman diagrams in Fig. 1\textbf{. }There is yet
another form of writing this, a Dyson equation,%
\begin{equation}
G_{{\small A,A}}^{R}(\varepsilon)=G_{{\small A,A}}^{oR}(\varepsilon
)+G_{{\small A,A}}^{R}(\varepsilon)\Sigma_{{\small A}}(\varepsilon
)G_{{\small A,A}}^{oR}(\varepsilon).\label{GAA-dyson}%
\end{equation}%
%TCIMACRO{\FRAME{ftbpFU}{7.3525cm}{5.5179cm}{0pt}{\Qcb{In the upper panel, the
%exact Green's function ( thick line) is represented as an infinite series of
%unperturbed Green's functions ( thin lines). Coupling matrix elements are
%dashed lines. The botton panel shows the self-consistent Dyson equation and
%the self-energy.}}{\Qlb{pastawskiFig1}}{pastawskiFig1.eps}%
%width 7.3525cm;  height 5.5179cm;  depth 0pt;  original-width 5.6759in;
%original-height 4.248in;  cropleft "0";  croptop "1";  cropright "1";
%cropbottom "0";  filename 'pastawskiFig1.eps';file-properties "XNPEU";}}}%
%BeginExpansion
\begin{figure}
[ptb]
\begin{center}
\includegraphics[
height=5.5179cm,
width=7.3525cm
]%
{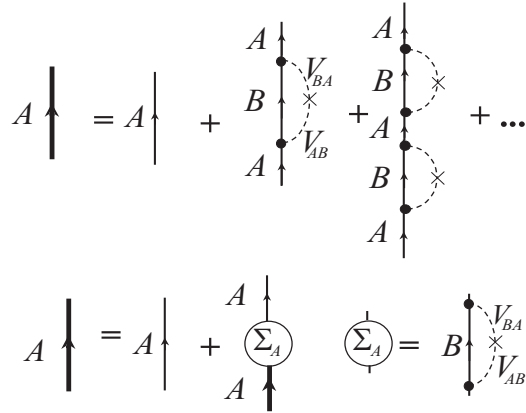}%
\caption{In the upper panel, the exact Green's function ( thick line) is
represented as an infinite series of unperturbed Green's functions ( thin
lines). Coupling matrix elements are dashed lines. The botton panel shows the
self-consistent Dyson equation and the self-energy.}%
\label{pastawskiFig1}%
\end{center}
\end{figure}
%EndExpansion
\textbf{\ }It is useful to note that all the above results, and most of what
follows, are also valid if $\left\vert A\right\rangle $ and $\left\vert
B\right\rangle $ denote whole subspaces. In that case, all the presented
equations and diagrams hold but with matrix elements transformed themselves
into matrices\cite{pastawski-MCF,Gascon}. We might choose not to deal
explicitly with an undesired subspace, for example the whole subspace
$\left\vert B\right\rangle ,$ and still get and effective Hamiltonian
restricted to the subspace $\left\vert A\right\rangle $ and also the exact
Green's function$.$

Usually, given an initial state, the dynamics is evaluated from eigen-energies
and eigenstates. Alternatively, it can be expressed in terms of Green's
functions. For example, the probability that a particle which was in the state
$\left\vert A\right\rangle $ at $t=0$ is found at state $\left\vert
B\right\rangle $ at a later time $t$ results:%
\begin{align}
P_{B,A}(t) &  =\left\vert \left\langle {\small B}\right\vert \exp
[-\mathrm{i}\hat{H}~t]\left\vert {\small A}\right\rangle \right\vert
^{2}\theta\lbrack t]\label{evolution-theta}\\
&  =\left\vert \lim_{\eta\rightarrow0^{+}}\int\frac{\mathrm{d}\varepsilon
}{2\pi\hbar}G_{{\small B,A}}^{R}(\varepsilon+i\eta)\exp[-i\varepsilon
t]\right\vert ^{2}\label{Evolution Retarded}\\
&  =\int\frac{\mathrm{d}\omega}{2\pi}\exp[-i\omega t]P_{{\small B,A}}%
(\omega)\\
&  =\int d\varepsilon P_{B,A}(\varepsilon,t),\label{energy-time-Wigner}%
\end{align}
with%
\begin{equation}
P_{{\small B,A}}(\omega)=\int\mathrm{d}\varepsilon\overset{P_{B,A}%
(\varepsilon,\omega)}{\overbrace{\tfrac{1}{2\pi\hbar}G_{{\small B,A}}%
^{R}(\varepsilon+\tfrac{1}{2}{\small \hbar\omega})G_{{\small A,B}}%
^{A}(\varepsilon-\tfrac{1}{2}{\small \hbar\omega})}}.\label{PAA(E,t)def}%
\end{equation}
The appearance of the function $\theta\lbrack t]$ in Eq. \ref{evolution-theta}
is consequence of the election of the sign of the imaginary part in the
retarded Green's function. The remaining two lines constitute alternatives for
writing the product of the independent integrals. The function $P_{B,A}%
(\varepsilon,t)$ (as well as its transform $P_{B,A}(\varepsilon,\omega)$) is
not an actual probability but a form of energy-time distribution function from
which a real probability can be obtained as a marginal distribution, i.e. by
integration of one of the variables.
%TCIMACRO{\FRAME{ftbpFU}{7.5259cm}{5.0259cm}{0pt}{\Qcb{(Color online)
%Energy-time distribution function for a two-level system (in units of $V$ and
%$\hbar/V$ respecively) The dark (yellow-red online) and clear (blue online)
%regions differ in sign. The formed stripes manifest the progressive decrease
%in the small structure's scale as function of time. }}{\Qlb{pastawskiFig2}%
%}{pastawskifig2.eps}{\special{ language "Scientific Word";  type "GRAPHIC";
%width 7.5259cm;  height 5.0259cm;  depth 0pt;  original-width 9.0817in;
%original-height 6.0228in;  cropleft "0";  croptop "1";  cropright "1";
%cropbottom "0";  filename 'pastawskiFig2.eps';file-properties "XNPEU";}}}%
%BeginExpansion
\begin{figure}
[ptb]
\begin{center}
\includegraphics[
height=5.0259cm,
width=7.5259cm
]%
{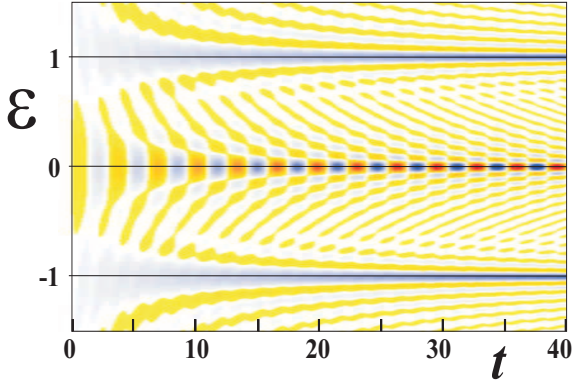}%
\caption{(Color online) Energy-time distribution function for a two-level
system (in units of $V$ and $\hbar/V$ respecively) The dark (yellow-red
online) and clear (blue online) regions differ in sign. The formed stripes
manifest the progressive decrease in the small structure's scale as function
of time. }%
\label{pastawskiFig2}%
\end{center}
\end{figure}
%EndExpansion
In more general problems, this energy-time distribution enabled
\cite{GLBE1,GLBE2} to consider time dependent statistical distribution
functions. For the particular case of equal energies $E_{{\small A}%
}=E_{{\small B}}=0$ and $V_{{\small A}B}=V$ with the superposition $\left\vert
A\right\rangle =\frac{1}{\sqrt{2}}\left(  \left\vert \bar{A}\right\rangle
+\left\vert \bar{B}\right\rangle \right)  $ as initial state:%
\begin{align}
P_{{\small A,A}}(\varepsilon,t) &  =\tfrac{V^{2}+\varepsilon\left(
V-2\varepsilon\right)  }{2\varepsilon\left(  V^{2}-\varepsilon^{2}\right)
}\sin\left[  2\left(  \varepsilon+V\right)  t\right]  \theta
(t)\label{PAA(E,t)explicit}\\
&  +\tfrac{V^{2}-\varepsilon\left(  V+2\varepsilon\right)  }{2\varepsilon
\left(  V^{2}-\varepsilon^{2}\right)  }\sin\left[  2\left(  \varepsilon
-V\right)  t\right]  \theta(t)\nonumber
\end{align}
This distribution oscillates as a function of each participant energy at a
rate which is determined by its distance to the eigenvalue (see Fig. 2). From
it, the Rabi oscillation is obtained as:%
\begin{equation}
P_{{\small A,A}}(t)=\int d\varepsilon P_{{\small A,A}}(\varepsilon,t)=\cos
^{2}(\tfrac{1}{2}\omega_{{\small AB}}t).\label{Paa}%
\end{equation}
Notice that while the result of the integral remains a simple oscillation, for
long times the integrand becomes an increasingly oscillatory function on the
energy variable. In a numerical integration, regions with too small structures
would contribute as pseudo-random amount to the integral making it numerically
unstable. It would be tempting to do an analogy with similar structures in the
standard momentum-position Wigner function suggested by Zurek
\cite{Zurek2003,CookPasJal}, and interpret this phenomenon as a manifestation
of the instability of this quantum superposition towards decoherence. In fact,
ideal Rabi oscillations contrast with experimental observations, such as Fig.
4-a of Ref. \cite{swapJCP2004}, where the environment is actually attenuating
the oscillation while the probability is conserved. Thus, our simple quantum
mechanical model should be extended to include some form of environmental interaction.

\section{The spectrum of a finite linear chain and continued fractions.}

We will represent the environment with our favorite model, the linear chain.
It not only represents a chain of spins interacting through a XY
interaction\cite{PUL96} but it is a reasonable model for polymers, quasi-one
dimensional crystals and metal wires. Even a crystal structure can be reduced
to a set of uncoupled linear chains. We start by adding a third state to our
two state system,%

\begin{equation}
\mathbf{H}=\left[
\begin{array}
[c]{ccc}%
E_{{\small 1}} & -V_{12} & 0\\
-V_{{\small 21}} & E_{{\small 2}} & -V_{{\small 23}}\\
0 & -V_{{\small 32}} & E_{{\small 3}}%
\end{array}
\right]  .\label{H3}%
\end{equation}
We start with $V_{{\small 12}}=0.$ Through the identification of the indices
$2\rightarrow A,$ and $3\rightarrow B$,~we use Eq.\ref{EeffA} eliminate state
$B$( i.e.$3$) so that $G_{{\small A,A}}^{R}(\varepsilon)$ $\rightarrow\bar
{G}_{{\small 2,2}}^{oR}(\varepsilon)$. Now we turn-on $V_{1,2}$ and identify
$1\rightarrow A$ and $2$ $\rightarrow B,$ and we repeat the elimination of $B$
to get:
\begin{equation}
G_{{\small 1,1}}^{R}(\varepsilon)=\frac{1}{\left[  G_{{\small 1,1}}%
^{oR}(\varepsilon)\right]  ^{-1}-V_{{\small 12}}\bar{G}_{{\small 2,2}}%
^{oR}(\varepsilon)V_{{\small 21}}}\label{G11(G22)}%
\end{equation}
We replace it and obtain a nested fraction:
\begin{equation}
G_{{\small 1,1}}^{R}(\varepsilon)=\frac{1}{\varepsilon-E_{1}-\underset
{\Sigma_{{\small 1}}}{\underbrace{V_{{\small 12}}\dfrac{1}{\varepsilon
-E_{2}-\underset{\Sigma_{{\small 2}}}{\underbrace{V_{{\small 23}}\dfrac
{1}{\varepsilon-E_{3}}V_{{\small 32}}}}}V_{{\small 21}}}}}\label{g(1,2,3)}%
\end{equation}
In the present context, the self-energy accounts for presence of states at the right.

Hamiltonian of Eq. (\ref{H3}) presents an interesting phenomenon. If
$V_{{\small 23}}\ll V_{{\small 12}}=V_{{\small AB}}$ the system $AB$ is well
defined and site $3$ can be seen as an \textquotedblleft
environment\textquotedblright\ weakly perturbing the system through
$V_{{\small SE}}=V_{{\small 23}}$. If we allow the parameters to switch to the
opposite regime $V_{{\small SE}}=V_{{\small 23}}\gg V_{{\small 12}%
}=V_{{\small AB}},$ state $B$ becomes \textquotedblleft
captured\textquotedblright\ by the environment and the state $A$ becomes
almost isolated. This can be seen as a\ form of the Quantum Zeno Effect
\cite{Misra-Sudarshan} caused by the internal degrees of freedom
\cite{Pastawski-Usaj,Pascazio}.

Since the procedure performed above was in fact a step of a renormalization
group algorithm\cite{Sokoloff-Jose,pastawski-MCF}, we can iterate it to get
the general continued-fraction that describes a chain with $N$ orbitals:%
\begin{equation}
\Sigma_{n}\left(  \varepsilon\right)  =V_{n,n+1}\dfrac{1}{\varepsilon
-E_{n}-\Sigma_{n+1}\left(  \varepsilon\right)  }V_{n+1,n}%
.\label{Sigma recurrence}%
\end{equation}
together with the termination condition.%
\begin{equation}
\Sigma_{{\small N}}\left(  \varepsilon\right)  \equiv0.\label{Sigma-end}%
\end{equation}

Hence, the Green's function, as the self-energy, is the ratio between two
polynomials. This yields the $N$ eigen-energies and eigenvalue weights of the
finite system. As predicted by Poincar\'{e} this produces many recurrences. A
particularly interesting dynamical recurrence is what we called \cite{PLU95}
the \textit{mesoscopic echo} which appears at the Heisenberg's time
$T_{ME}=\hbar/\bar{\Delta}$ where $\bar{\Delta}$ is the mean level spacing.
Signatures of this phenomenon where experimentally observed in C\'{o}rdoba
\cite{PUL96} and confirmed in Zurich by the group of Richard R. Ernst as can
be seen in Fig. 3-B of Ref. \cite{Ernst-mesos-Echo}.

\section{The semi-infinite ordered chain}

When the chain of lattice spacing $a$ is ordered ($E_{n}\equiv0,~~V_{n,n+1}%
\equiv V$) and infinite there is no termination condition as Eq.
\ref{Sigma-end}. Instead, all sites \textquotedblleft see\textquotedblright%
\ the same environment at their right. Hence, the equation that is now
equivalent to the Bloch theorem is
\begin{equation}
\Sigma_{n}\left(  \varepsilon\right)  \equiv\Sigma_{n+1}\left(  \varepsilon
\right)  =\Sigma\left(  \varepsilon\right)  ,
\end{equation}
from which:%

\begin{equation}
\Sigma\left(  \varepsilon\right)  =\dfrac{V^{2}}{\varepsilon-\Sigma\left(
\varepsilon\right)  }.\label{Sigma-Dyson}%
\end{equation}
The surprise is that in the region where \textit{there are} real eigenvalues,
the solution is complex
\begin{equation}
\Sigma\left(  \varepsilon\right)  =\Delta\left(  \varepsilon\right)
-\mathrm{i}\Gamma\left(  \varepsilon\right)  ,\label{Sig=D+iF}%
\end{equation}
the energy shift is a piece-like function:
\begin{equation}
\Delta\left(  \varepsilon\right)  =\left\{
\begin{array}
[c]{l}%
\underset{}{\dfrac{\varepsilon}{{\small 2}}}-\sqrt{\left(  \dfrac{\varepsilon
}{{\small 2}}\right)  ^{2}-V^{2}}\,\,\,\mathrm{for\,\,\,\,}\varepsilon
>2\left\vert V\right\vert ,\\
\dfrac{\varepsilon}{{\small 2}}%
\,\,\,\,\ \,\,\,\,\,\,\,\,\,\,\,\,\,\,\,\,\,\ \,\,\,\,\,\,\,\,\,\,\mathrm{for\,\,\,\,\,\,}%
\left\vert \varepsilon\right\vert \leq+2\left\vert V\right\vert ,\\
\dfrac{\varepsilon}{{\small 2}}+\sqrt{\left(  \dfrac{\varepsilon}{{\small 2}%
}\right)  ^{2}-V^{2}}\,\,\mathrm{for\,\,\,}\varepsilon<-2\left\vert
V\right\vert .
\end{array}
\right.  \label{(Delta)}%
\end{equation}
while the group velocity, $\Gamma=\hbar v_{\varepsilon}/a$, results
\begin{equation}
\Gamma\left(  \varepsilon\right)  =\left\{
\begin{array}
[c]{c}%
0\,\,\ \,\,\,\,\,\,\ \,\,\,\,\ \,\,\,\,\,\ \,\,\,\,\,\,\ \mathrm{for\,\,\,\,}%
\varepsilon>2\left\vert V\right\vert ,\\
\sqrt{V^{2}-\left(  \dfrac{\varepsilon}{{\small 2}}\right)  ^{2}%
}\,\,\,\,\mathrm{for\,\,\,\,\,}\left\vert \varepsilon\right\vert
\leq+2\left\vert V\right\vert ,\\
0\,\,\,\,\,\,\ \,\,\,\,\,\,\ \,\,\,\,\,\,\,\,\,\ \,\,\ \mathrm{for\,\,\,\,}%
\varepsilon<-2\left\vert V\right\vert .
\end{array}
\right.  \label{F}%
\end{equation}
The sign of the square root is consistent with the analytical properties
described above, while the real part goes to zero as $\lim_{\varepsilon
\rightarrow\pm\infty}\Delta\left(  \varepsilon\right)  =0$ which means that
the spectrum of the linear chain remains bounded after the interaction has
been turned-on. The consistency of these solutions can be checked through the
convergence of the self-energies in chains of increasing lengths. This
expresses the \textit{Quantum Thermodynamic Limit}:%
\begin{align}
-\Gamma\left(  \varepsilon\right)   &  =\lim_{\eta\rightarrow0^{+}}%
\lim_{N\rightarrow\infty}\operatorname{Im}\Sigma_{1}^{{}}\left(
\varepsilon+\mathrm{i}\eta\right)  \label{Gamma lim n N}\\
&  \neq\lim_{N\rightarrow\infty}\lim_{\eta\rightarrow0^{+}}\operatorname{Im}%
\Sigma_{1}^{{}}\left(  \varepsilon+\mathrm{i}\eta\right)  \underset
{\mathrm{a.e.\varepsilon}}{\equiv}0\label{Gamma lim N n}%
\end{align}
$\mathrm{a.e.}\varepsilon$ means for almost every $\varepsilon\mathrm{,}$ i.e.
except for a set whose probability measure is zero. The non-triviality of this
limit is manifested in the fact that it is non-uniform.

\section{The\ Fermi Golden Rule as a Quantum Thermodynamic Limit}

In the above discussion we obtained an effective energy with an imaginary
component. It actually means that perturbation theory does not converge. The
unperturbed eigenstate is so far from the new eigenstates that their scalar
product vanishes. In the dynamics, this should manifest as a progressive decay
where the Poincar\'{e} recurrences no longer appear. This means that the
probability escapes towards the semi-infinite chain. For the homogeneous
linear chain this involves a power law decay according to the law
$P_{1,1}(t)\simeq\left(  Vt\right)  ^{-1}$. A particularly interesting case
occurs when at the end (surface) of this semi-infinite chain we add an orbital
(or atom) with energy $E_{0}$ and interaction $V_{0}\ll V.$ This adatom model,
is a particular case of the Friedrichs model. One knows that this situation
leads to a typical exponential decay described by the Fermi Golden Rule (FGR).
However, a deeper analysis shows that the exact rate of decay differs from
that in the FGR. The new rate, $\Gamma_{0}/\hbar$, arises from a Self
Consistent Fermi Golden Rule \cite{SC-FGR}. It is the imaginary part at the
exact pole $\varepsilon_{r}-\mathrm{i}\Gamma_{o}$ of the Green's function:%
\begin{equation}
\varepsilon_{r}-\mathrm{i}\Gamma_{0}=E_{0}+\frac{V_{0}^{2}}{V^{2}}%
~\Sigma(\varepsilon_{r}-\mathrm{i}\Gamma_{o}) \label{Er pole}%
\end{equation}
which can be obtained analytically or by iteration.

One should not forget that a quantum decay starts always quadratically, in
this case with a time scale $\hbar/V_{0}.$ It only starts looking exponential
after a time $t_{S}$. This is a short time scale,
\begin{equation}
t_{S}=\hbar\pi~\bar{N}_{1}(\varepsilon_{r}), \label{tS}%
\end{equation}
when the escape from the surface site towards the rest of the chain prevents
the return and hence stops giving an appreciable contribution to the survival.
Here, $\bar{N}_{1}(\varepsilon_{r})$ is the LDoS at the surface site in
absence of the adatom. At times longer than,%
\begin{equation}
t_{R}=\alpha\frac{\Gamma_{0}}{\hbar}\ln\left[  \beta\frac{B}{\Gamma_{0}%
}\right]  , \label{tR}%
\end{equation}
the return amplitude, determined by the high order processes that has already
escaped but remains in the neighborhood, starts being comparable to the pure
survival amplitude. From then on, decay becomes a power law $\left[
\Gamma(\varepsilon_{r})t\right]  ^{-3/2}$. Here, $B=4V$ is the bandwidth and
$\alpha$,$\beta\gtrsim1$ are constants that depend on the van Hove
singularities of $\bar{N}_{1}(\varepsilon_{r})$ and other details of the
model. At $t_{R}$ a striking destructive interference between the pure
survival amplitude and the return amplitude may occur. In quantum systems,
this \textquotedblleft survival collapse\textquotedblright\ \cite{SC-FGR} has
yet to be observed.

In summary, the validity of the FGR is restricted by memory effects to times
between $t_{R}$ and $t_{S}$. The standard FGR holds in the wide band limit
$\Gamma_{0}/B\rightarrow0$ which also implies that $V_{0}\bar{N}%
_{1}(\varepsilon_{r})\rightarrow0.$ It is only in this condition, valid in a
quite broad variety of situations, that one can forget the quantum memory
effects of a finite bandwidth and replace both $\Delta\left(  \varepsilon
\right)  -\mathrm{i}\Gamma\left(  \varepsilon\right)  $ by $\Delta
-\mathrm{i}\Gamma$ independent of $\varepsilon.$ The environment behaves as a
Markovian process and we refer to them as the \textquotedblleft broad band
approximation\textquotedblright\ or \textquotedblleft fast fluctuations
approximation\textquotedblright. One should be careful, however, interpreting
this as an \textquotedblleft irreversible\textquotedblright\ behavior
\cite{Prigogine}. Actual irreversibility is consequence of an instability that
manifests when one attempts to revert the dynamics through a \textquotedblleft
Loschmidt daemon\textquotedblright, $\hat{H}\rightarrow-\hat{H}$
\cite{JalPas}. One should note that an imaginary part by itself does not
ensures irreversibility as long a one can change the sign of the whole
Hamiltonian. The instability of this procedure can be tested and quantified
through the Loschmidt echo (or fidelity) in spin dynamics
\cite{JCP98,PhysicaA}, confined atoms \cite{DAVIDSON-echo} and microwaves in
cavities\cite{stockmann}. See also Ref. \cite{TimeReversalMirror} for a
completely different approach to achieve time-reversal.

The physical meaning of the imaginary part we introduced at the beginning is
now evident: it represents the weak interaction with an environment. In such
situation, $\Delta$ does not contribute much to the dependence on
$\varepsilon$ and one includes it by shifting the energies. This approximation
would give a steady decay of the Rabi oscillation as is indeed typical of many
experimental setups. See for example the Fig. 4-a in Ref \cite{swapJCP2004}.
However, one might wonder how to return\ the probability lost in this decay.
In fact in presence of two identical linear chains connected to states $A$ and
$B$, one would find probability \cite{levstein-swapp},
\begin{align}
\bar{P}_{{\small A,A}}(t) &  =P_{{\small A,A}}^{{}}(t)\exp\left[  -2\Gamma
t/\hbar\right]  \label{PABdecay}\\
&  =\cos^{2}(2\omega_{{\small AB}}t)\exp\left[  -t/\tau\right]  ,\mathrm{with}%
~\tau=\hbar/2\Gamma.\nonumber
\end{align}
Clearly, this describes the evolution of polarization tunneling between two
nuclei shown in Fig. 4.b of Ref. \cite{swapJCP2004}. In this case, the
probability (polarization) is not conserved but it decays according to the
FGR. While this could be correct in some physical situations, the description
of a situation closer to Fig. 4.a, where probability is conserved,\ remained a challenge.

\section{The Generalized Landauer-B\"{u}ttiker Equation}

The imaginary energy has been a puzzle for everyone using Green's functions
and regularizing its poles. Sometimes, as in the electron-phonon processes, an
explicit form for this imaginary energy is evaluated through the FGR. Even the
transport equations, as the Kubo formula, rely on some natural broadening
which enables the computation but produces local non-conservation of currents.
The answer was given by D'Amato and Pastawski \cite{DaPa} who, extending an
idea of B\"{u}ttiker\cite{Buttiker}, realized that the escape to an
environment is equivalent to saying that, at each time, a fraction of the
system occupation escapes to the chain which could act as a voltmeter. As an
actual voltmeter, however it should not extract net particles from the system,
so it returns a particle for each one collected. This can be expressed
\cite{GLBE1} in terms of the Landauer description of transport which now
accounts for time dependences and decoherent process in the form of a
Generalized Landauer-B\"{u}ttiker Equation (GLBE). Hence, for every process of
\textquotedblleft escape\textquotedblright\ from the coherent beam due to the
interaction with the environment, a fresh incoherent particle must be
reinjected into the system as expressed in Eq. (3.7) of Ref. \cite{GLBE1}.
This physical picture, finds its formal justification when the
system-environment interactions are local and the environment spectrum is so
broad that it becomes instantaneous and energy independent. In this case, the
Keldysh quantum field theory formalism, expressed in its integral form
\cite{Keldysh2}, reduces to the GLBE \cite{GLBE2,Pastawski-Medina} represented
in Fig. 3.%
%TCIMACRO{\FRAME{ftbpFU}{7.5634cm}{3.9996cm}{0pt}{\Qcb{Diagrams for the density
%propagator from $A$ to $B$ as dictated by Generalized Landauer-B\"{u}ttiker
%Equation. Horizontal lines are single particle Green\U{b4}s functions dressed
%by the environment. Shadowed vertices are the self-consistent density
%propagators. The vertical double dashed lines represent the reinjection
%processes. The last collision occurs at site $n$.}}{\Qlb{pastawskiFig3}%
%}{pastawskifig3.eps}{\special{ language "Scientific Word";  type "GRAPHIC";
%width 7.5634cm;  height 3.9996cm;  depth 0pt;  original-width 7.2552in;
%original-height 7.2709in;  cropleft "0";  croptop "1";  cropright "1";
%cropbottom "0.4770";  filename 'pastawskiFig3.eps';file-properties "XNPEU";}%
%}}%
%BeginExpansion
\begin{figure}
[ptb]
\begin{center}
\includegraphics[
trim=0.000000in 3.468219in 0.000000in 0.000000in,
height=3.9996cm,
width=7.5634cm
]%
{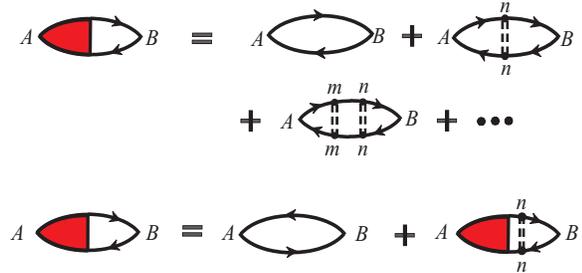}%
\caption{Diagrams for the density propagator from $A$ to $B$ as dictated by
Generalized Landauer-B\"{u}ttiker Equation. Horizontal lines are single
particle Green\'{}s functions dressed by the environment. Shadowed vertices
are the self-consistent density propagators. The vertical double dashed lines
represent the reinjection processes. The last collision occurs at site $n$.}%
\label{pastawskiFig3}%
\end{center}
\end{figure}
%EndExpansion

We consider a degenerate two-level system where, besides tunneling, each state
suffers the action of a complex self-energy, such as that of the linear chain
described in the previous section. This results in an homogeneous interaction
with the environment. The natural parameter regulating the effectiveness of
the system-environment is $g=\omega_{AB}\Gamma/\hbar.$ For this two-level
system the GLBE becomes:%
\begin{align}
\widetilde{P}_{A,A}(t)  &  =\bar{P}_{A,A}(t)\label{AB-GLBE}\\
&  +%
%TCIMACRO{\dsum \limits_{n=A,B}}%
%BeginExpansion
{\displaystyle\sum\limits_{n=A,B}}
%EndExpansion
\int_{0}^{t_{i}}\bar{P}_{A,n}(t-t_{i})\frac{\mathrm{d}t_{i}}{\tau}%
\widetilde{P}_{n,A}(t_{i}),\nonumber
\end{align}
and equivalent equations for the components $BA$, $AB$ and $BB.$ This is a
Volterra's type equation. This is a Dyson equation (much as Eq.
\ref{GAA-dyson}) for a density, i.e. a two-particle Green's function, and is
also known as a Bethe-Salpeter equation.\ The first term describes the
probability of coherent propagation from the initial to the final state which
decays due to interactions with the environment. The kernel of this equation
is precisely $\bar{P}_{A,n}(t-t_{i}),$ the two-particle propagator. Since
$\mathrm{d}t_{i}/\tau$ is the probability of having the last interaction with
the environment at the time interval $\mathrm{d}t_{i}$ around $t_{i}$. The
solution of the homogeneous GLBE can be obtained by Fourier transformation
\cite{Danieli-SSC07} $\widetilde{P}_{B,A}(\omega)$ and decays toward the
equilibrium $\widetilde{P}_{B,A}(t)\rightarrow\frac{1}{2}$. One notable thing
is that the first term in the right has poles in the complex $\omega$-plane
that correspond to the difference of energies and do not present any form of
non- analyticity. The self consistent solution $\widetilde{P}_{B,A}(\omega)$
has more information. In fact, the poles of $\delta\widetilde{P}_{B,A}%
(\omega)=\widetilde{P}_{B,A}(\omega)-\frac{1}{2}\delta(\omega)$ are precisely
at
\begin{equation}
\omega^{\pm}-\mathrm{i}\Gamma=\pm\sqrt{\left[  \omega_{AB}\right]  ^{2}%
-\Gamma^{2}}-\mathrm{i}\Gamma
\end{equation}
The trajectories in the complex plane are shown in Fig. 4-b. The important
feature is that the real part of the poles (Fig. 5-a) collapses at $0$ for a
critical value $g_{c}=1$ and from this point they split in two terms of null
real part. One of them decreases with environment interaction whereas the
other decreases. It is the later that controls the long time behavior.%
\begin{equation}
\delta\widetilde{P}_{A,A}(t)=\widetilde{P}_{A,A}(t)-\tfrac{1}{2}=a_{0}%
\cos\left[  (\omega+\mathrm{i}\Gamma)t+\phi\right]  . \label{Paa(t)eff}%
\end{equation}
Here $P_{A,A}^{\mathrm{eq.}}\equiv\frac{1}{2}$ is the equilibrium occupation
while $a_{0}^{2}=\left[  4\omega^{2}\tau^{2}+1\right]  /\left(  16\omega
^{2}\tau^{2}\right)  $ and $\phi=\arctan\left[  1/2\omega\tau\right]  $
warrant the initial cuadratic decay.%
%TCIMACRO{\FRAME{ftbpFU}{7.5119cm}{4.0699cm}{0pt}{\Qcb{a) Paths of poles of a
%single particle Green's function, e.g. $G_{AA}^{R}(\varepsilon)$, when an
%homogeneous decay $\Gamma$is increased. They move parallel to the imaginary
%axis. b) Paths of poles of the observable $\delta\widetilde{P}_{AA}(\omega)$
%(a two-particle self-consistent Green\U{b4}s funcion) when $\Gamma$ increases.
%The symmetric frequencies collapse at the center where a branching occurs. One
%mode becames long life while the other has a short life time.}}%
%{\Qlb{pastawskiFig4}}{pastawskiFig4.eps}%
%width 7.5119cm;  height 4.0699cm;  depth 0pt;  original-width 5.7949in;
%original-height 3.117in;  cropleft "0";  croptop "1";  cropright "1";
%cropbottom "0";  filename 'pastawskiFig4.eps';file-properties "XNPEU";}}}%
%BeginExpansion
\begin{figure}
[ptb]
\begin{center}
\includegraphics[
height=4.0699cm,
width=7.5119cm
]%
{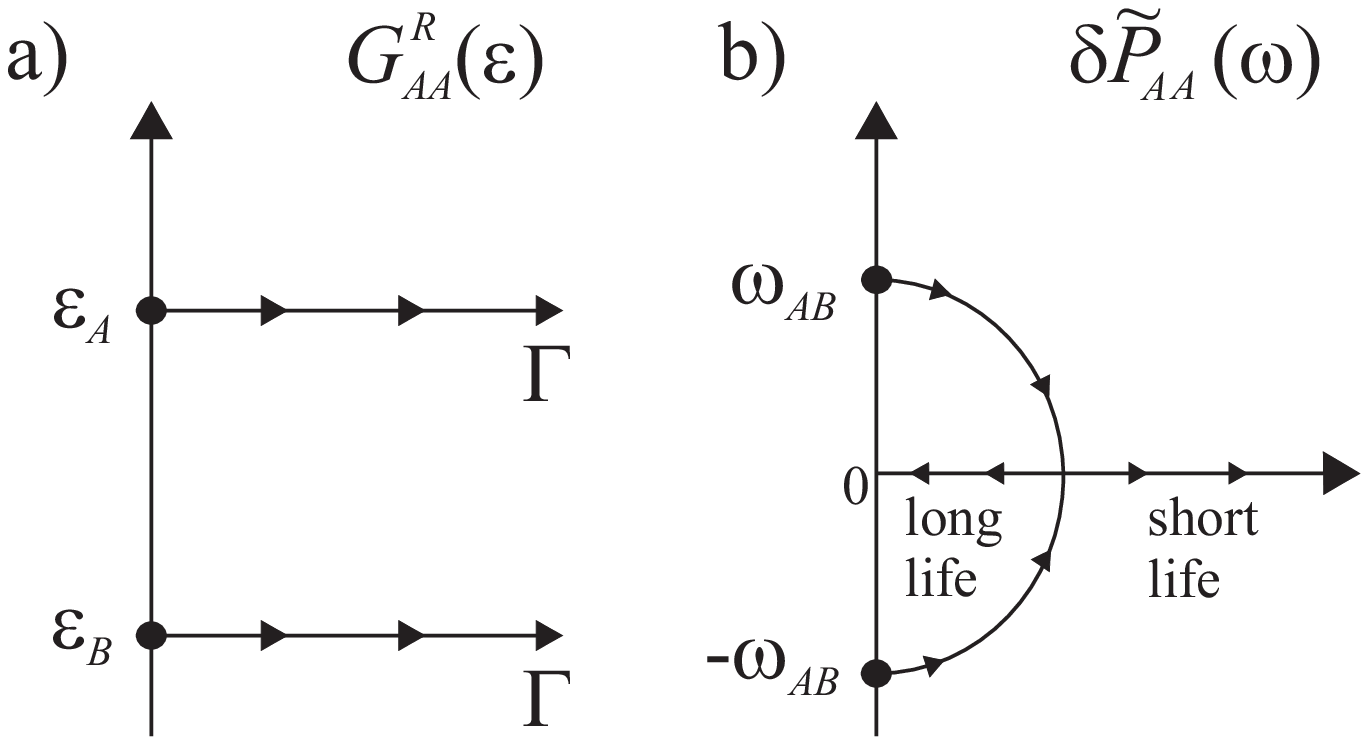}%
\caption{a) Paths of poles of a single particle Green's function, e.g.
$G_{AA}^{R}(\varepsilon)$, when an homogeneous decay $\Gamma$is increased.
They move parallel to the imaginary axis. b) Paths of poles of the observable
$\delta\widetilde{P}_{AA}(\omega)$ (a two-particle self-consistent Green\'{}s
funcion) when $\Gamma$ increases. The symmetric frequencies collapse at the
center where a branching occurs. One mode becames long life while the other
has a short life time.}%
\label{pastawskiFig4}%
\end{center}
\end{figure}
%EndExpansion

The described behavior has been experimentally observed in heterogeneous
polarization transfer, see Fig. 7 of Ref. \cite{JCP98}, but overlooked since
the early theory for this experiment \cite{Ernst-exp-swapp} did not contain
the transition. Recent experimental studies \cite{Alvarez-exp-swapp} show the
divergence of the period $2\pi/\omega$ at a critical ratio $\hbar\omega
_{AB}/\Gamma=1.$ Conceptually, the transition is from an isolated system that
is weakly perturbed to a state in which the effect of the environment is no
longer perturbative.~The system may be no longer well defined as discussed
with reference to Eq. \ref{H3}. This would be a dynamical Quantum Zeno Effect
\cite{Pastawski-Usaj,Pascazio}. While the limiting cases were somehow
expected, it was by no means obvious that this change could be critical. The
non-analyticity was enabled by the infinite degrees of freedom of the
environment in the proper quantum thermodynamic limit and the self-consistent
nature of Eq. \ref{AB-GLBE}.

The excess density $\delta\widetilde{P}_{A,A}(t)$ behaves exactly as the
amplitude $x(t)$ of a damped classical oscillator which undergoes a transition
to an overdamped regime. Indeed, considering a damped harmonic oscillator of
mass $m$ and natural frequency $\omega_{o}^{{}}$, the relaxation rate as a
function of the friction coefficient, $\Gamma,$ follows precisely the trace of
Fig 5-b: The rate increases with friction until a critical value when it
starts to decrease inversely proportional to the friction coefficient $\Gamma
$. This is, of course, a non-analytic critical behavior. Where does it come
from? From the imaginary self-energy correction that shifts the natural
frequency $\omega_{o}^{{}}$ in the oscillator's dynamical susceptibility
$\chi(\omega)=$ $-m^{-1}/\left[  \omega^{2}-(\omega_{o}^{2}-\mathrm{i}%
\omega\Gamma)\right]  .$ The damped Newton's equation is not a fundamental law
but it is written on phenomenological grounds. However, the inclusion of
$\Gamma$ can be justified, within statistical mechanics, by including the
action of a Brownian bath \cite{Ingold}. Recently, we obtained a simpler
demonstration \cite{Calvo-osciladores} using as environment a chain of
oscillators whose $N$ degrees of freedom are considered by taking the
thermodynamic limit of $N\rightarrow\infty$ precisely in the same way as
described above in the context of the FGR. It is interesting to note that
while $2\omega_{o}^{{}}/\Gamma\gg1$ corresponds to the standard oscillation.
In a similar way, in the quantum case $2\omega_{o}^{{}}/\Gamma\gg1,$ the
system is well defined and the environment is a small perturbation. In
contrast in the regime controlled by friction, $2\omega_{o}^{{}}/\Gamma\ll1,$
the inertia term can be completely neglected.

It is clear that most of the qualitative features of the spectral properties
described above are valid for other linear systems (provided that there is a
thermodynamic limit) and hence are ubiquitous in Nature. In magnetic
resonance, a phenomenon known as exchange narrowing, has long been described
\cite{KUBO} and clearly observed \cite{Calvo-exchange-narrowing}. However, its
explanation requires either Brownian fluctuations or the use of Markov chains
with imaginary probabilities...!\cite{Anderson-exchange}.

\section{Phase Transitions as Paradigm Shifts}

In the previous sections we have touched upon issues such as complex energies,
imaginary probabilities, irreversibility, recurrences, decoherence,
non-analytic observables, etc., all of them generating strong polemics. In
consequence, some epistemological comments are pertinent.

One of the central statements of ancient Physics was Aristotle's dictum that
everything that moves is moved by something else. More precisely, Aristotle
says that the velocity of a moving object is directly proportional to the
force and inversely proportional to the resistance, i.e. $\dot{x}=F/\Gamma$.
In the absence of a proximate force, the body would come to rest immediately.
Obviously, a difficulty found in the Aristotelian view is the justification of
why a projectile keeps moving through the air. The logic of the explanation is
not as clean as the central statement: a projectile would owe its continuing
motion to the force of eddies or vibrations in the surrounding medium, a
phenomenon known as antiperistasis. This was formalized later on by the
scholastics \cite{Buridam} who proposed that motion was maintained by some
property of the body, the \textit{impetus,} which once set in motion, would
impart the force keeping the movement. Buridan's impetus has the same
consequence, but very different justification, than the modern concept of
momentum \cite{Feyerabend}.%
%TCIMACRO{\FRAME{ftbpFU}{7.2986cm}{7.8047cm}{0pt}{\Qcb{a)The frequency of a
%two-level system (Fig. 4b) collapses at zero for a critical $\Gamma$. b) the
%decoherence rate as function of $\Gamma$. This also represents a relaxation
%rate in a damped harmonic oscillator as function of friction strength. The
%botton left point is ideal frictionless Hamiltonian mechanics or Newton's
%paradigm. The right side is the realm of Aritotle's paradigm where inertia
%becomes negligible. }}{\Qlb{pastawskiFig5}}{pastawskiFig5.eps}%
%width 7.2986cm;  height 7.8047cm;  depth 0pt;  original-width 3.6124in;
%original-height 3.8688in;  cropleft "0";  croptop "1";  cropright "1";
%cropbottom "0";  filename 'pastawskiFig5.eps';file-properties "XNPEU";}}}%
%BeginExpansion
\begin{figure}
[ptb]
\begin{center}
\includegraphics[
height=7.8047cm,
width=7.2986cm
]%
{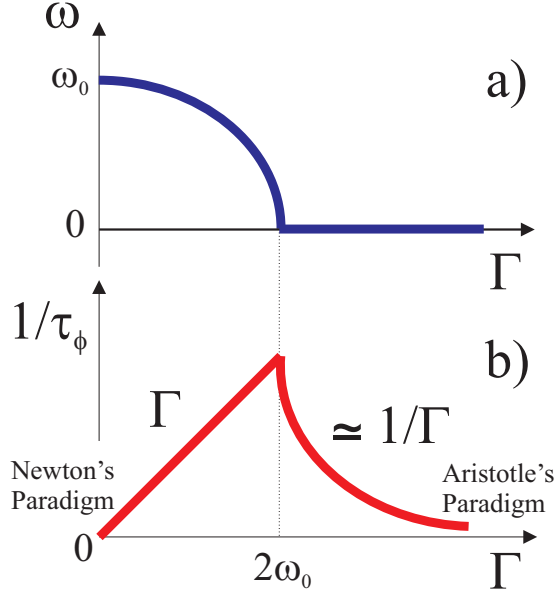}%
\caption{a)The frequency of a two-level system (Fig. 4b) collapses at zero for
a critical $\Gamma$. b) the decoherence rate as function of $\Gamma$. This
also represents a relaxation rate in a damped harmonic oscillator as function
of friction strength. The botton left point is ideal frictionless Hamiltonian
mechanics or Newton's paradigm. The right side is the realm of Aritotle's
paradigm where inertia becomes negligible. }%
\label{pastawskiFig5}%
\end{center}
\end{figure}
%EndExpansion

Physics seemed a quite solid construction until the experiments and intuition
of Galileo and analytical skills of Newton started to build much of our
current knowledge. In this new formulation, the \textit{inertia} is no longer
a correction but the \textit{fundamental principle}. Friction appears now as a
subsidiary and phenomenological effect needed to fit the natural phenomena to
the ideal scheme. Indeed its justification remained for a long time alien to
Hamiltonian mechanics. One had to wait for the appearance of Boltzmann's
statistical mechanics and the work of Smoluchowski and Einstein to have a
place in the theory building (for a simple Hamiltonian model justifying
friction see Ref. \cite{Calvo-osciladores}). In any case, Aristotelian and
Newtonian views, were so completely irreconcilable that Thomas Kuhn
\cite{Kuhn-what-are} concluded that they were indeed different views of
Nature. He coined the term \textit{paradigm shift} to describe a change in the
basic assumptions within the ruling theory of science. According to Kuhn,
science progress relies not only on a linear accumulation of new knowledge as
sustained by Karl Popper but, more fundamentally, on periodic revolutions in
which the nature of scientific inquiry within a particular field is abruptly
transformed \cite{Kuhn-struct}. Rival paradigms are said to be
\textit{incommensurable} because it is not possible to understand one paradigm
through the conceptual framework and terminology of another rival paradigm.

What seems disturbing to some scientists, is the possibility that no ultimate
truth underlies this confrontation between paradigms \cite{Weinberg-on-Kuhn}.
Is it possible to synthesize these extreme behaviors into a single framework?
Our answer is yes, because incommensurability involves comparing states at
different sides of a phase transition. Indeed, Aristotle's paradigm is placed
at the extreme right side of Fig. 5-b where the inertia's contribution to an
equation of motion is completely neglected. The impetus corrections allows one
to move somehow to the left. The contrasting Newton's paradigm, $\ddot{x}%
=F/m$, is placed at the extreme left, in the origin of Fig. 5-b. From that
ideal point one could conceive adding friction as a correction. Consider a
mass placed in a bowl where friction can be varied. Would anyone experimenting
in one of those extremes conceive, without completing the experiment of the
other regime, what the response at the other edge would be? The answer is a
clear no, as the non-analytic function does not allow a natural extrapolation.
Indeed, it was not until Gauss popularized the concept and interpretation of
Euler's complex numbers that both regimes fitted into a single description.
Even with that tool, numerous discussions with students and colleagues
convinced me that intuition fails lamentably at the non-analytic point. The
same occurs when one discusses problems which involve the non-homogeneity of
the limits, which indeed is at the root of the microscopic description of
friction. Many other controversies in Physics have a resolution within this
framework: we have already advocated that the Loschmidt vs. Boltzmann
controversy is a consequence of the non-uniformity of the limits for an
imperfect time reversal experiment \cite{PhysicaA}. Each argument results
valid in a different approach to the limiting case (see Fig. 6 in Ref.
\cite{CookPasJal}). The Zermelo/Poincar\'{e}-Boltzmann controversy is another
consequence of different forms of taking the thermodynamic limit.

More recently, in the quantum framework, the localized-extended transition
owes its origin to the fact that strong disorder induces a non-uniformity of
the limits respect to ensemble average,%
\begin{align}
\bar{\Gamma}(\varepsilon)  &  =\lim_{\eta\rightarrow0^{+}}\left\langle
\lim_{N\rightarrow\infty}\operatorname{Im}\Sigma(\varepsilon+\mathrm{i}%
\eta)\right\rangle _{%
%TCIMACRO{\QTATOP{\QTR{rm}{ens.}}{\QTR{rm}{ave.}}}%
%BeginExpansion
\genfrac{}{}{0pt}{1}{\mathrm{ens.}}{\mathrm{ave.}}%
%EndExpansion
}\label{Gamma n ave}\\
&  \neq\left\langle \lim_{\eta\rightarrow0^{+}}\lim_{N\rightarrow\infty
}\operatorname{Im}\Sigma(\varepsilon+\mathrm{i}\eta)\right\rangle _{%
%TCIMACRO{\QTATOP{\QTR{rm}{ens.}}{\QTR{rm}{ave.}}}%
%BeginExpansion
\genfrac{}{}{0pt}{1}{\mathrm{ens.}}{\mathrm{ave.}}%
%EndExpansion
}\underset{\mathrm{a.e.}\varepsilon}{\equiv}0. \label{Gamma ave eta}%
\end{align}

This inequality and the last equality were proved and tested numerically in
Ref. \cite{Pasta-Slutzky}. They show that in the localized regime the spectrum
is pure-point. Not recognizing it led to contradictory results for about two
decades \cite{errorANDERSON,Rodrigues}. Also the coarse grain average has
subtle properties of non-uniformity respect to the thermodynamic limit which
need further exploration \cite{Foa-Pasta-Medina}.

There are other smaller paradigm shifts in condensed matter physics, which
resulted somehow less conflictive, produced by the need to explain quantum
phase transitions. We can mention superconductivity (from current carried by
single electrons to Cooper's pairs), localization and mesoscopic transport
(which shifted from Kubo's view where dissipation occurs inside the sample to
that of Landauer, where it occurs at the external reservoirs) and the Integer
Quantum Hall (where the standard vision of bulk current yields to
B\"{u}ttiker's edge current).

Finally, I feel the obligation to mention another phase transition which
should not be much different from that discussed above: the transition from
static friction to dynamical friction. In that case, ordinates in Fig. 5-b
describe the friction force as a function of the applied force. The abrupt
fall of the last at a critical force describes the transition to the almost
constant value of the dynamical friction. In fact, the non-analytic jump from
static friction to dynamical friction is so unexpected and counter-intuitive
that no other phase transition seems to have a bigger deathly tall in
\textquotedblleft accidents\textquotedblright\ on the road, at work or even at
home. It seems to me that it is a most urgent challenge to devise an
educational strategy capable to develop, in the general public and physicists
alike, an intuition on this phenomenon. On the physical side, friction has
only recently been reintroduced as a fundamental problem \cite{frictionKRIM}.
Its formulation relies on models having a close connection to issues discussed
above \cite{modelsSOKOLOFF}. This is still another phase transition that opens
new questions not only for basic physics but, even more importantly, also to
social and cognitive sciences.

\section{Acknowledgements}

It is a pleasure to acknowledge the physicists from whom I received my
education: A. L\'{o}pez D\'{a}valos, J. F. Weisz, M. C. G. Passeggi, P. A. Lee
and B. L. Altshuler (I hope they recognize any of their seeds flourishing
through my work). I am also indebted to my life-long collaborator and
companion P. R. Levstein and to my students J. L. D'Amato, G. Usaj, J. A.
Gasc\'{o}n, F.\ M. Cucchietti, L. E. F. Fo\`{a} Torres, E. P. Danieli, G. A.
\'{A}lvarez, E. Rufeil Fiori, H. L. Calvo, A. Dente and G. Ludue\~{n}a because
of what I learned while teaching them. The hospitality of Abdus Salam ICTP
enabled many beneficial discussions. This work was financed by grants from
Fundaci\'{o}n Antorchas, CONICET and SeCyT-UNC.

\end{document}